\documentclass[12PT]{iopart}

\usepackage{graphicx}

\newcommand{\sqrtsNN}{\mbox{$\sqrt{s_{_{\mathrm{NN}}}}$}}
\newcommand{\dAu}{\textit{d}+Au}

\newcommand{\AuAu}{Au+Au}

\newcommand{\pp}{\mbox{\textit{p}+\textit{p}}}

\newcommand{\PT}{\mbox{$p_T$}}

\newcommand{\gevc}{\mbox{${\mathrm{GeV/}}c$}}

\newcommand{\RAA}{\mbox{$R_{AA}$}}

\newcommand{\npart}{\mbox{$N_{\mathrm{part}}$}}

\begin{document}
\title[Open heavy flavor production at RHIC]
{Open heavy flavor production at RHIC}

\author{A A P Suaide}

\address{Instituto de F\'{i}sica da Universidade de S\~ao Paulo, Departamento de F\'{i}sica
Nuclear, Caixa Postal 66318, 05315-970 S\~ao Paulo, SP, Brazil}

\ead{suaide@if.usp.br}

\begin{abstract}
The study of heavy flavor production in relativistic heavy ion
collisions is an extreme experimental challenge but provides
important information on the properties of the Quark-Gluon Plasma
(QGP) created in \AuAu\ collisions at RHIC. Heavy-quarks are
believed to be produced in the initial stages of the collision, and
are essential on the understanding of parton energy loss in the
dense medium created in such environment. Moreover, heavy-quarks can
help to investigate fundamental properties of QCD in elementary p+p
collisions. In this work we review recent results on heavy flavor
production and their interaction with the hot and dense medium at
RHIC.
\end{abstract}


High energy nuclear collisions at the Relativistic Heavy Ion
Collider (RHIC) \cite{Roser2002} have opened a new domain in the
exploration of strongly interacting matter at very high energy
density. High temperatures and densities are generated in central
nuclear collisions, creating the conditions in which a phase of
deconfined quarks and gluons exists (QGP) \cite{Blaziot1999}. The
measurements at RHIC in conjunction with theoretical calculations
suggest that a dense, equilibrated system has been generated in the
collision with properties similar to that of an ideal hydrodynamic
fluid. The strong suppression phenomena observed for high-\PT\
hadrons \cite{STARPRL89,STARPRL90,STARPRL91} suggest that the system
early in its evolution is extremely dense and dissipative.

Heavy quark (charm and bottom) measurements further expand the
knowledge about the matter produced in nuclear collisions at RHIC.
Because of their large masses, their production can be calculated by
pertubative QCD (pQCD) \cite{pQCD}. In particular, comparative
measurements in \pp, \dAu\ and \AuAu\ are sensitive to the initial
state gluon densities in these systems \cite{Muller1992}. In \AuAu\
collisions, medium effects such as heavy quark energy loss can be
studied through comparison of \PT\ distributions of bottom and charm
production with those observed for inclusive hadrons. The
suppression of small angle gluon radiation (dead cone effect)
predicted for heavy quarks would decrease the amount of energy loss
\cite{Dok2001,Zhang2004} though gluon emission and, therefore, the
suppression of heavy quark mesons at high-\PT\ is expected to be
smaller than that one observed for light quark hadrons at RHIC. In
this case, smaller energy loss allows heavy quarks to probe deeper
into the medium \cite{SIMON}. It also opens new possibilities to
investigate other interaction mechanisms such as collisional energy
loss \cite{elastic, Djordjevic:2005, Armesto, Wicks:2005, Hess:2005}
and in medium fragmentation \cite{vitev}. Moreover, measuring open
charm and bottom production at RHIC provides essential reference
data for studies of color screening via quarkonium suppression
\cite{Abreu}.

\section{Open heavy flavor measurements at RHIC}

The study of heavy flavors in relativistic nuclear collisions
follows two different approaches: (i) the direct reconstruction of
heavy flavored mesons and (ii) the study of semi-leptonic decays of
such mesons.  In this section we describe these methods, used by
STAR and PHENIX, and discuss their advantages and limitations.

\begin{figure}[t]
\begin{minipage}{8cm}
\includegraphics[width=1\textwidth]{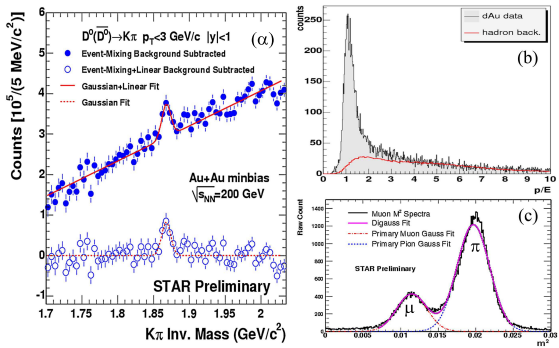}
\end{minipage}
\begin{minipage}{7cm}
\hspace{-6pc}
\parbox[h]{7cm}{
\label{Fig_reco} \caption{Heavy flavor identification methods used
at RHIC. (a) $D^0\rightarrow K^-\pi^+$ invariant mass spectrum after
event mixing subtraction in \AuAu\ collisions from STAR. (b) $p/E$
spectrum used for electron identification in \dAu\ collisions from
STAR. Similar method is also used by PHENIX. (c) Time of Flight mass
spectrum used for muon identification in STAR.} }
\end{minipage}
\end{figure}

\subsection{Direct reconstruction of D-mesons}

Direct reconstruction of heavy-flavor mesons is being performed by
the STAR collaboration using the decay channel $D^0\rightarrow K^-
\pi^+$ (and c.c.) with branching ratio of 3.83\% in \dAu\ and \AuAu\
collisions. Because of the small branching ratio and the lack of
dedicated detector triggers, the direct reconstruction of $D$-mesons
requires the analysis of large amount of data. The available
statistics limits the study of such mesons to the low-\PT\ region
($\PT<3$ GeV/c). Kaons and pions are identified using the STAR Time
Projection Chamber (TPC) $dE/dx$. The resulting invariant mass
spectrum of kaon-pion pairs contains a substantial amount of
background from random combinatorics that can be subtracted using
mixed event methods. Details of the analysis can be found in Ref
\cite{STAR_ToFe}. Figure \ref{Fig_reco}-a shows an invariant mass
distribution after event mixing subtraction where a clear $D^0$ peak
can be seen.

Despite the limitation on \PT\ and statistics, direct reconstruction
of $D$-mesons is the cleanest probe to investigate heavy quarks in
relativistic nuclear collisions.

\subsection{Semi-leptonic decays of D and B mesons}

Direct reconstruction of $D$-mesons can be done only by STAR in a
limited \PT\ range and still difficult to perform with high
efficiency in high multiplicity events. In this case semi-leptonic
decays of such mesons (Ex: $D^{0}\rightarrow e^{+} + K^{-}+\nu$)
provide a more efficient measurement of charm and bottom production.
At RHIC two methods are utilized to measure heavy flavor production
via semileptonic decays: (i) identification of electrons from $D$
and $B$-meson decays and (ii) identification of muons from $D$-meson
decays.

The analysis of non-photonic electrons (the excess of electrons
after subtracting all possible sources of background, such as photon
conversions and Dalitz decays) is a technique used by STAR and
PHENIX. Electron identification in STAR is done using the $dE/dx$
and momentum ($p$) information from the TPC and the Time of Flight
(ToF) data for low-\PT\ ($\PT < 4-5$ GeV/c) electrons
\cite{STAR_ToFe} or energy ($E$) and shower shape in the
electromagnetic calorimeter (EMC) for high-\PT\ ($\PT> 1.5$ GeV/c)
electrons \cite{STAR_HPTe} (see Figure \ref{Fig_reco}-b).

Electron identification in PHENIX \cite{PHENIX_HPTeAA,PHENIX_HPTePP}
is largely based on the Ring Imaging Cherenkov detector (RICH) in
conjunction with a highly granular EMC. The momentum is derived from
drift and pad chambers.

A major difficulty in the electron analysis for both experiments is
the fact that there are many sources of electrons other than
semi-leptonic decays of heavy-flavor mesons. The main sources of
background are photon conversion in the detector material (less
significant in PHENIX due to the reduced amount of material when
compared to STAR) and $\pi^0$ and $\eta$ Dalitz decays. Other
sources of background, such as $\omega$, $\phi$ and $\rho$ decays
are also taken into account, although their contribution is small
compared to the sources mentioned above. These background sources
are commonly called photonic electrons.

Background subtraction in the PHENIX experiment is performed by two
different methods: (i) The converter method, in which a well defined
amount of material is added to the detector to increase the number
of background electrons. By comparing the electron spectra with and
without the converter it is possible to measure directly a
significant part of the background. (ii) The cocktail method, with
which PHENIX measures the spectra of the main background sources
(photons, $\pi^0$ and $\eta$). Both methods agree to each other very
well.

Due to its the large acceptance and tracking efficiency, STAR can
directly reconstruct a considerable fraction of the photonic
background by performing invariant mass reconstruction of $e^+e^-$
pairs with high efficiency. For photon conversion, $\pi^0$ and
$\eta$ Dalitz decays the invariant mass spectrum shows a peak near
zero mass. Other background sources are evaluated by simulations and
account for a very small fraction of the total background in this
case.

Muon identification at low-\PT\ ($\PT < 0.25$ GeV/c) plays an
important role because this \PT\ range imposes a significant
constrain on the measurement of charm cross section \cite{MUONS}.
This measurement is being done by STAR using $dE/dx$ and momentum
information from the TPC and mass reconstruction from the ToF
detector. A clear muon signal can be reconstructed, as seen in
Figure \ref{Fig_reco}-c. Background from $\pi$ and $K$ decays can be
subtracted by looking at the DCA (distance of closest approach)
between the muon and the primary vertex of the collision.

\section{Charm production at RHIC}

\begin{figure} [t]
\begin{center}
\includegraphics[width=0.8\textwidth]{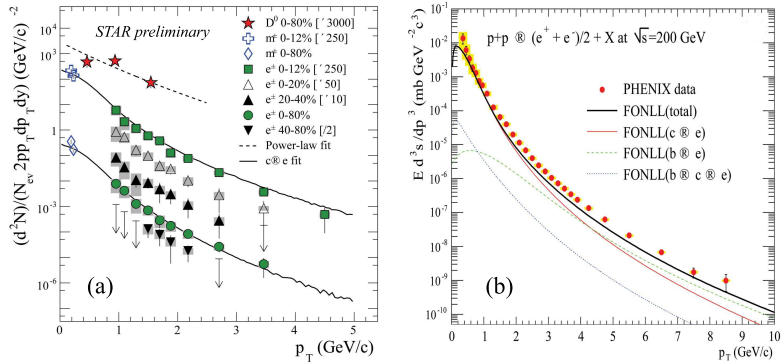}
\end{center}
\label{Fig_XS} \vspace{-0.5cm} \caption{Data used by STAR and PHENIX
to study total charm production at RHIC. (a) $D^0$, $\mu$ and
electron data used in a combined fit from STAR. (b) electron data
used by PHENIX. Curves are FONLL prediction used by PHENIX to
extrapolate the electron spectra to $\PT=0$.}
\end{figure}

The determination of the total charm cross section via non-photonic
electrons requires precise measurements down to very low-\PT. This
is an experimental challenge: the lower the \PT, the higher the
amount of photonic electrons that contaminates the measurement,
resulting in large systematic uncertainties. In order to overcome
this limitation STAR performs three independent measurements: direct
reconstruction of D-mesons, $\mu$ and $e^{\pm}$. The total charm
cross section is obtained from a combined fit of these measurements
\cite{MUONS} as illustrated in Figure \ref{Fig_XS}-a. PHENIX, on the
other hand, uses its $e^{\pm}$ measurements to extract total cross
sections \cite{PHENIX_HPTePP}. To overcome the large background at
low-\PT\ PHENIX is improving analysis techniques in order to reduce
systematic uncertainties due to photonic background. This allowed a
decrease in the minimum \PT\ that can be measured by PHENIX from
$\sim 0.8$ in the earlier datasets to $\sim 0.4$ GeV/c (Figure
\ref{Fig_XS}-b).

\begin{figure} [t]
\begin{center}
\includegraphics[width=0.45\textwidth]{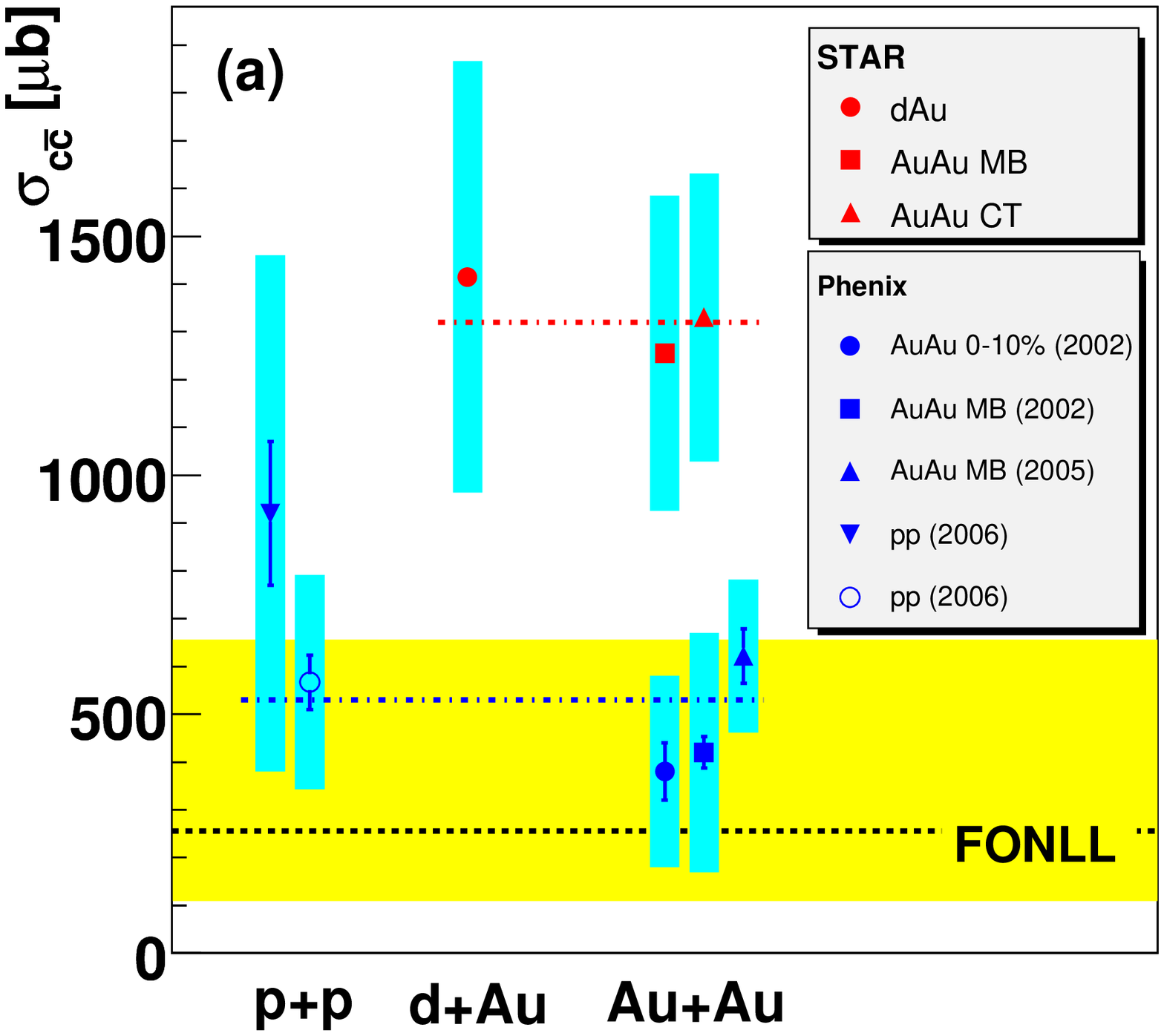}
\includegraphics[width=0.45\textwidth]{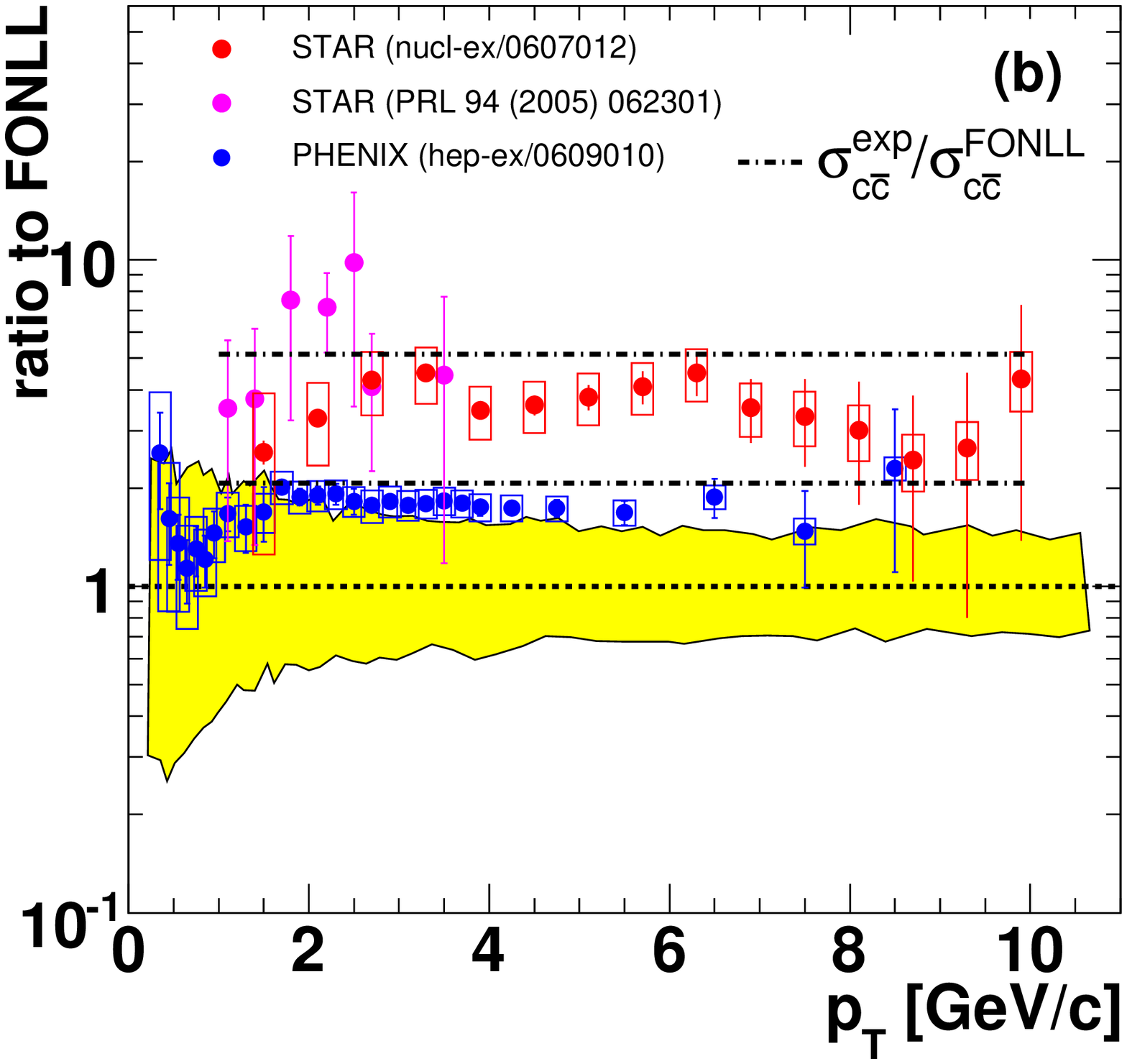}
\end{center}
\vspace{-0.5cm}
\caption{(a) Overview of the experimental charm
cross section at RHIC from \pp\ to central \AuAu collisions. (b)
ratio of measured non-photonic electron yield to FONLL pQCD
calculations for \pp\ collisions.} \label{FIGSUMMARY}
\end{figure}

Figure \ref{FIGSUMMARY}-a summarizes all charm cross section
measurements  from \pp\ to central \AuAu\ collisions at
$\sqrtsNN=200$ GeV at RHIC
\cite{STAR_ToFe,PHENIX_HPTeAA,PHENIX_HPTePP,MUONS,PHENIX_SIGMA}.
Shown is the total cross section per binary collision. Figure
\ref{FIGSUMMARY}-a makes evident a factor of $\sim2$ between STAR
and PHENIX, this difference being larger than the combined
systematic uncertainties in the case of central \AuAu\ collisions.
The dashed line in this figure depicts the average prediction from
FONLL calculations \cite{FONLL} for charm production at RHIC
energies and the yellow band its uncertainty determined by
independent variation of quark masses, renormalization and
factorization scales. The dash-dotted lines correspond to the
average values for both experiments. Despite the differences between
experiments the results, for each experiment, suggest that charm
production follows a binary scaling from \pp\ to central \AuAu\
collisions. This is a strong indication that charm are predominantly
produced in the early stages of the collision evolution and other
processes, such as thermal production in the QGP, are not
significant.

Figure \ref{FIGSUMMARY}-b shows the ratio of STAR \cite{STAR_HPTe}
and PHENIX \cite{PHENIX_HPTePP} measurements to FONLL calculations
for high-\PT\ electrons in \pp\ collisions at RHIC. The dash-dotted
lines correspond to the ratio of the experimental to FONLL cross
section. This makes evident that the factor of $\sim2$ discrepancy
in the cross section extends up to large \PT\ values. Note that,
despite this normalization discrepancy FONLL describes the shape of
the measured spectra well in both cases, suggesting that the
differences between the measurements may be related to an
experimental normalization effect.

Future RHIC measurements will address this experimental discrepancy.
STAR is planning a run  without its inner tracking detectors (SVT
and SSD) in the next years thus reducing the amount of photon
conversion and will address in detail some of the systematic
uncertainties in the background removal of the electron
measurements. Added to that, both STAR and PHENIX are developing
detector upgrades providing drastic improvements in secondary vertex
reconstruction which will allow the use of displaced vertex
techniques to direct measure $D$ and $B$ mesons with high precision
and efficiency.

\section{Modifications of heavy flavors in the medium}

Putting aside the discrepancies in the absolute cross sections
measurements between STAR and PHENIX we now investigate how heavy
quarks interact with the QGP formed in \AuAu\ collisions at RHIC.
Over the last few years RHIC is providing interesting information on
how the partons behave in such hot and dense medium. The study of
flavor dependence of these interactions will further expand our
knowledge about the properties of the nuclear matter under such
extreme conditions.

\subsection{Energy loss of heavy quarks}

\begin{figure} [t]
\begin{center}
\hspace{-0.60cm}
\includegraphics[width=0.36\textwidth]{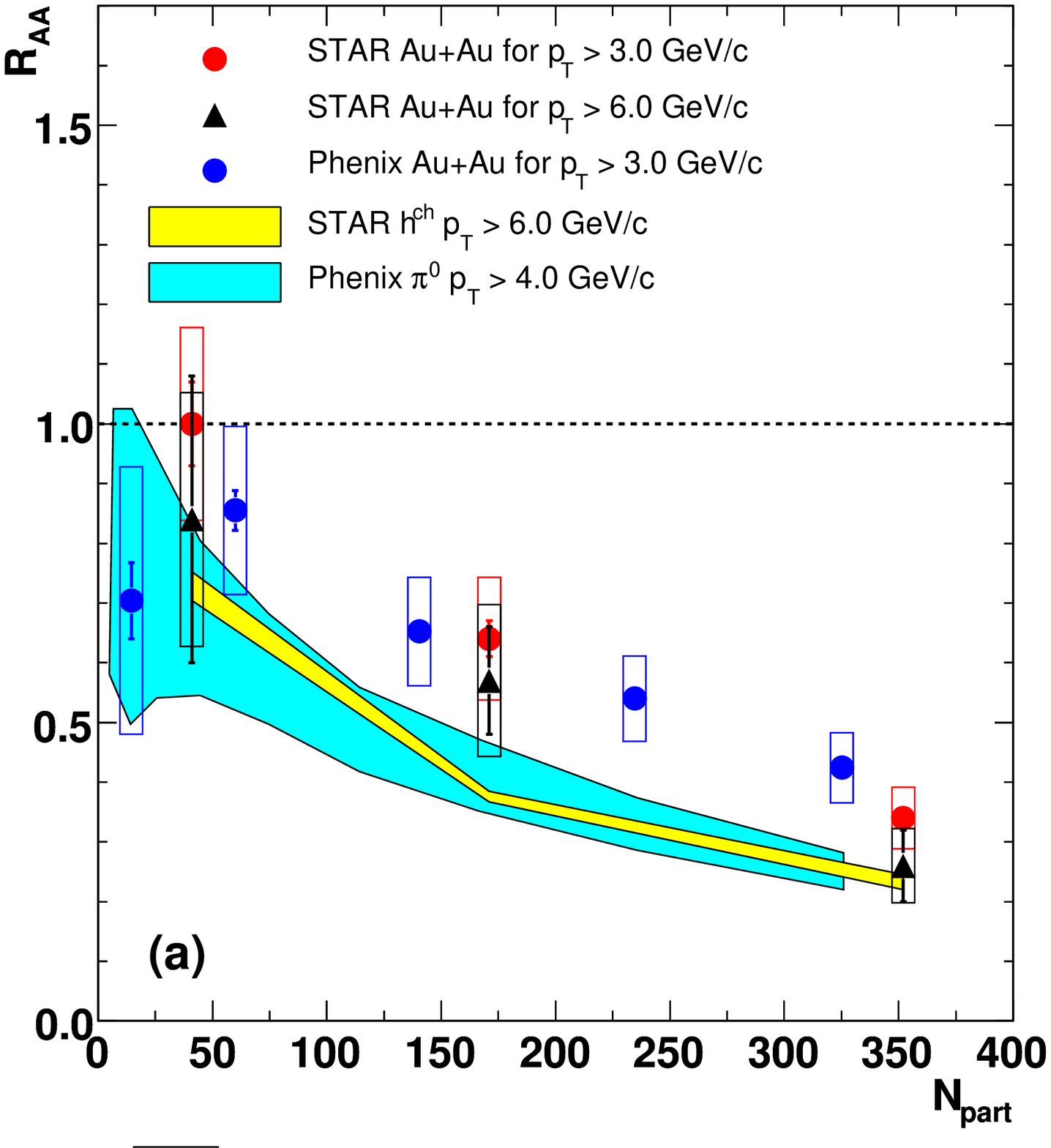}
\hspace{-0.65cm}
\includegraphics[width=0.36\textwidth]{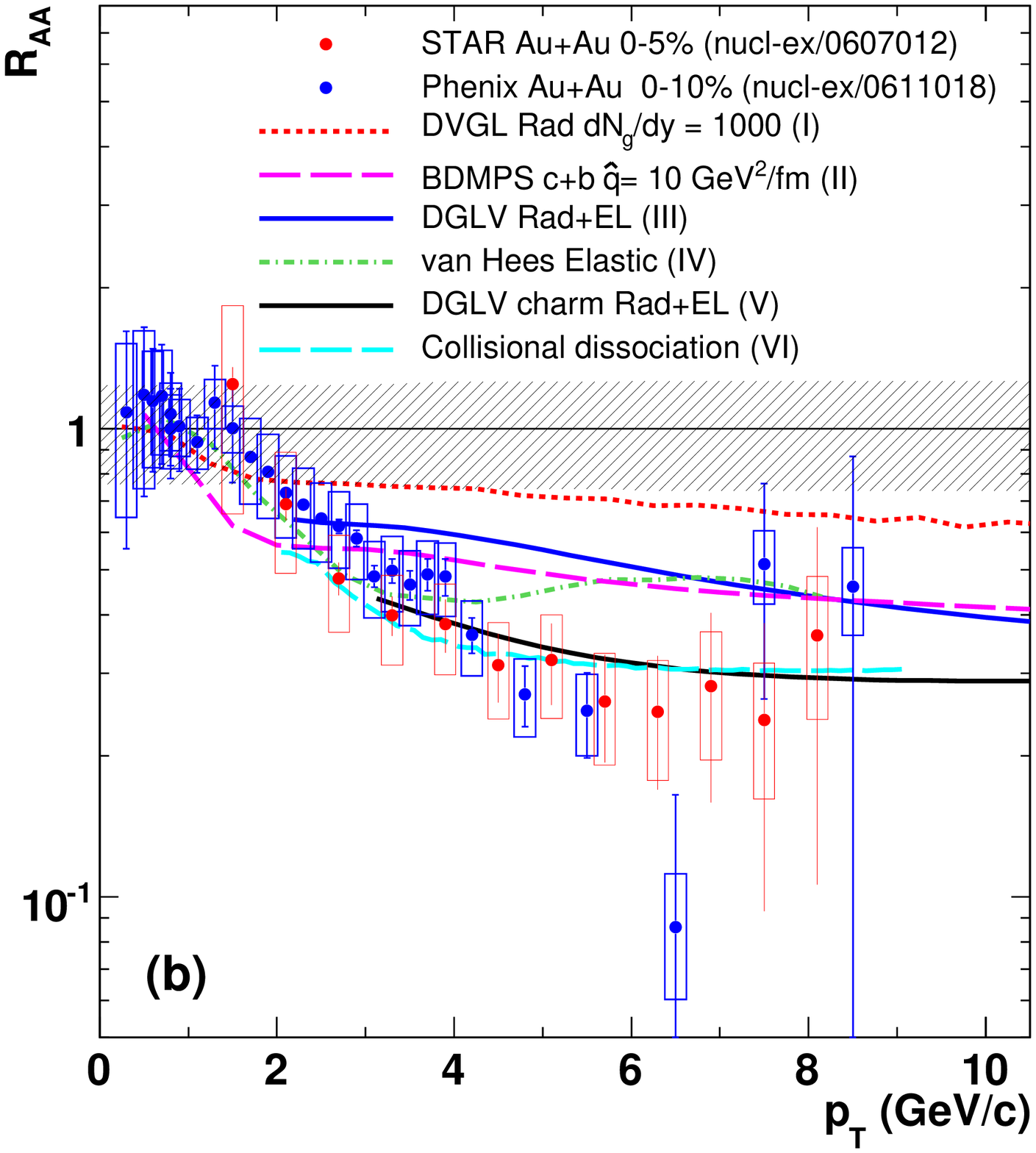}
\hspace{-0.65cm}
\includegraphics[width=0.36\textwidth]{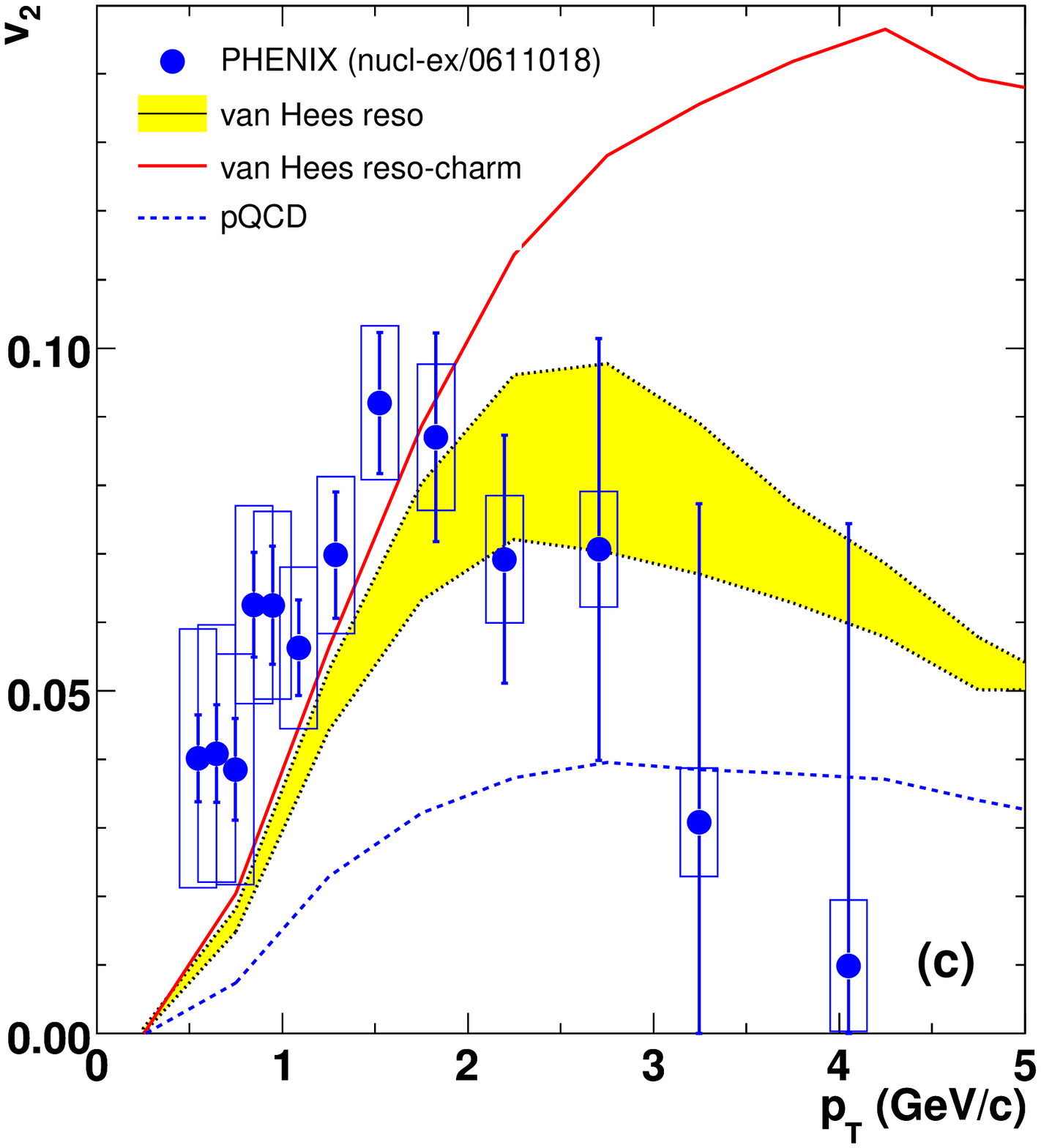}
\end{center}
\vspace{-0.5cm} \caption{Modification of non-photonic electrons in
\AuAu\ collisions at RHIC. (a) \RAA\ as a function of \npart\; (b)
\RAA\ as a function of \PT\ for central collisions (see text). (c)
$v_2$ as a function of \PT\ for minimum bias collisions.}
\label{Fig_RAA}
\end{figure}

Nuclear effects in non-photonic electron production are measured
through the comparison of spectra from \dAu\ and \AuAu\ collisions
to the equivalent spectrum in \pp: the relevant quantity is the
ratio $\RAA(\PT) = (dN_{AA}/d\PT)/(T_{AA} \times dN_{pp}/d\PT)$,
where $dN_{AA}/d\PT$ is the differential yield in \AuAu\ (\dAu) and
$dN_{pp}/d\PT$ the corresponding yield in \pp\ collisions. $T_{AA}$
is the nuclear overlap integral, derived from Glauber calculations
\cite{GLAUBER}. In the absence of nuclear effects, such as
shadowing, Cronin effect, or gluon saturation, hard processes are
expected to scale with the number of binary collisions and hence,
\RAA(\PT) = 1. Figure \ref{Fig_RAA}-a shows the average \RAA\ for
high-\PT\ non-photonic electrons as a function of \npart\ for STAR
and PHENIX data. Despite the cross section discrepancies between
STAR and PHENIX, \RAA\ results are consistent with each other. \RAA\
for non-photonic electrons shows an increasing suppression from
peripheral to central \AuAu\ collisions, indicating an unexpectedly
energy loss of heavy quarks in the medium. This suppression is
similar to the one observed for light-quark hadrons, indicated by
the shaded area in the figure. For the 5\% most central collisions,
non-photonic electron production for $\PT > 6$ \gevc\ is suppressed
by a factor $\sim 5$.

The suppression of non-photonic electrons in central \AuAu\
collisions can be, to some degree, explained in terms of radiative
energy loss in the medium. Figure \ref{Fig_RAA}-b also shows
different theoretical predictions for the non-photonic electron
suppression in central \AuAu\ collisions considering different
energy loss mechanisms. Curves (I) and (II) correspond to the
expected heavy quark suppression when the energy loss mechanism is
induced gluon radiation, considering electrons from $D$ and $B$
mesons decays. Curve (I) corresponds to the average value of DVGL
radiative energy loss calculation in which the medium gluon density
is $dN_{g}/dy=1000$ \cite{Djordjevic:2005}. In this case, the
radiative energy loss does not account for the observed suppression,
although the uncertainties, both in the data and theory, are large.
On the other hand, Curve (II) \cite{Armesto} shows a larger
suppression than Curve (I). In this case, however, the sensitivity
of \RAA\ to the time averaged transport coefficient $\hat{q}$
becomes smaller as it increases. In fact, the variation in \RAA\ for
$4 < \hat{q}$ (GeV$^{2}$/fm) $<14$ at $p_{T}> 3$ GeV/c is only $\sim
0.15$ \cite{Armesto}. This saturation of the suppression for high
values of $\hat{q}$ can be attributed to the highly opacity of the
medium, biasing the particle production towards its surface.
Although gluon radiation is still expected to be a significant
energy loss mechanism other processes may become important to
describe the suppression observed in central \AuAu\ collisions.
Curve (III) is an average prediction for electrons from heavy-quark
mesons decays and includes both DVGL radiative and elastic energy
loss, as well as jet path length fluctuations \cite{Wicks:2005}. In
contrast with light quarks, the elastic energy loss is comparable to
the radiative for heavy quarks \cite{elastic} and the effect on
\RAA\ is significant. In fact, theoretical predictions for elastic
rescattering of partons in the medium that can create resonant D and
B meson states via quark coalescence \cite{Hess:2005} in the medium
can lead to a significant suppression at moderate \PT\, as seen in
Curve (IV). In this case, the amount of suppression depends on the
resonance widths. Other processes, such as in-medium fragmentation
\cite{vitev}, shown in Curve (VI) may also contribute to the
observed suppression in central \AuAu\ collisions. It is clear that
the full understanding of the energy loss mechanisms is a
fundamental milestone for the characterization of the medium
properties. Heavy flavors provide an important tool to investigate
these mechanisms.

\subsection{Elliptic flow of heavy flavors}

It has been argued that the matter created in heavy-ion collisions
at RHIC is sufficiently hot and dense that charm quarks might
thermalize in the medium \cite{THERM}. The most promising method to
study this hypothesis is the measurement of charm elliptic flow.
Large elliptic flow for $D$-mesons would imply in a large number of
interactions, enough to suggest thermalization.

Recent PHENIX results \cite{PHENIX_HPTeAA} for non-photonic electron
$v_2$, shown in Figure \ref{Fig_RAA}-c, suggest that charm quarks
carry a substantial amount of $v_2$. Non-photonic electron $v_2$ at
low-\PT\ is compatible with rescattering+resonances calculations
suggesting a reduction in thermalization times of heavy quarks in
the medium, when compared to pQCD scattering \cite{Hess:2005}. At
high-\PT\ we notice a tendency of dropping in the non-photonic
electron $v_2$, although the statistical uncertainties are very
large. This could be an indication of increasing dominance of
electrons from bottom productions since bottom is not assumed to
show a pronounced $v_2$. More precise data and the independent
measurement of $D$-meson $v_2$ is fundamental to understand $v_2$ in
this \PT\ region.

\section{The next step: D and B mesons contributions to
semi-leptonic measurements}

Generally, all current models overpredict \RAA\ at high-\PT. It is
important to note that in all calculations charm quarks are
substantially more quenched than bottom quarks. Curve (V) in Figure
\ref{Fig_RAA}-b, which is based only on electrons from $D$ decays
describes the data best. It is the dominance of electrons from $B$
decays for $\PT > 4-5$ \gevc\ that pushes the predicted \RAA\ to
higher values. All theoretical calculations use the relative
contribution of $D$ and $B$ as predicted by pQCD calculations
\cite{FONLL}. The understanding of the observed suppression as well
the observed $v_2$ of non-photonic electrons require independent
measurements of $D$ and $B$ mesons at high-\PT.

\begin{figure}[t]
\begin{minipage}{8cm}
\hspace{+1cm}
\includegraphics[width=0.67\textwidth]{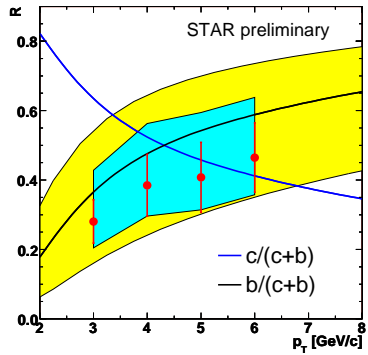}
\end{minipage}
\begin{minipage}{7cm}
\hspace{-8pc}
\parbox[h]{7cm}{
\caption{Preliminary results on the relative contribution of $B$
meson decays to the non-photonic electron spectra in \pp\ collisions
at RHIC \cite{XIAOYAN}.} \label{FIGCORREL}}
\end{minipage}
\end{figure}

$e-h$ $\Delta\phi$ correlations in \pp\ collisions is a promising
tool to measure this relative contribution. In general, heavy
flavors have harder fragmentation functions than gluons and light
quarks, making the near side correlation ($\Delta\phi\sim0$) more
sensitive to the decay kinematics. Figure \ref{FIGCORREL} shows
preliminary data from STAR on the relative contribution of charm and
bottom to the non-photonic electron data. The relative contribution
is obtained from the fit of the near side correlation with
predictions for $e-h$ correlations for $D$ and $B$ mesons from
Pythia. Experimental uncertainties contain the ones from photonic
background removal (dominant part) and the choice of fit method.
Details of this analysis can be found in Ref \cite{XIAOYAN}. The
lines show FONLL predictions for charm and bottom. The yellow band
reflects the uncertainty in the calculation. Within errors, the
experimental data is compatible with FONLL calculations. This
result, if confirmed, would imply that bottom may be more suppressed
in central \AuAu\ collisions than predicted by current energy loss
calculations. In order to further investigate the relative
contribution of charm and bottom to non-photonic electrons one may
look for $e-D^0$-meson correlations at $\Delta\phi \sim \pi$ in \pp\
collisions. Because of the decay channels of $D$ and $B$ mesons the
study of the charge dependence ($e^--D^0$ and $e^+-D^0$) of this
correlations can help on the separation of electrons from
$c\overline{c}$ and $b\overline{b}$ pairs.

\section{Final remarks}

RHIC is providing challenging data on heavy flavor production and
its interplay with the medium. Recent results show that this
interaction with the medium is stronger than what we expected
resulting in a large energy loss in central \AuAu\ collisions. In
order to take the next step it is imperative to independently
measure nuclear modification factors and $v_2$ of $D$ and $B$
mesons. This will be possible in the near future when detector
upgrades are available. Detailed and systematic measurements are
required to address existing discrepancies and are fundamental to
our understanding of heavy flavor production at RHIC.

The author thanks the STAR and PHENIX collaborations for providing
preliminary data used in this work. The author also thanks R.
Averbeck, M. Djordjevic, J. Dunlop, H. van Hees, X. Lin, R. Rapp, T.
Ullrich, I. Vitev, R. Vogt and  Z. Xu and Y.F. Zhang for valuable
discussion on this subject over the last months. This work is
partially supported by FAPESP and CNPq of Brazil.

\end{document}